\documentclass[aps,prb,reprint,superscriptaddress,longbibliography,floatfix]{revtex4-2}

\usepackage{amsmath,amssymb,graphicx,bm}
\usepackage{xcolor}
\usepackage[colorlinks=true,urlcolor=blue,linkcolor=blue,citecolor=blue,bookmarks=false]{hyperref}

\begin{document}

\title{Magnetic properties of the noncentrosymmetric tetragonal antiferromagnet EuPtSi$_{3}$}

\author{A. Bauer}
 \email{andreas.bauer@ph.tum.de}
 \affiliation{Physik-Department, Technische Universit\"{a}t M\"{u}nchen, D-85748 Garching, Germany}
 \affiliation{Zentrum f\"ur QuantumEngineering (ZQE), Technische Universit\"at M\"unchen, D-85748 Garching, Germany}

\author{A. Senyshyn}
 \affiliation{Heinz Maier-Leibnitz Zentrum (MLZ), Technische Universit\"{a}t M\"{u}nchen, D-85748 Garching, Germany} 

\author{R. Bozhanova}
\affiliation{Physik-Department, Technische Universit\"{a}t M\"{u}nchen, D-85748 Garching, Germany}

\author{W. Simeth}
\altaffiliation[Present address: ]{Paul Scherrer Institut (PSI), CH-5232 Villigen, Switzerland}
\affiliation{Physik-Department, Technische Universit\"{a}t M\"{u}nchen, D-85748 Garching, Germany}

\author{C. Franz}
 \affiliation{Physik-Department, Technische Universit\"{a}t M\"{u}nchen, D-85748 Garching, Germany}
 \affiliation{J\"ulich Centre for Neutron Science (JCNS) at Heinz Maier-Leibnitz Zentrum (MLZ), D-85748 Garching, Germany}
 
\author{S. Gottlieb-Sch\"{o}nmeyer}
 \affiliation{Physik-Department, Technische Universit\"{a}t M\"{u}nchen, D-85748 Garching, Germany}

\author{M. Meven}
 \affiliation{J\"ulich Centre for Neutron Science (JCNS) at Heinz Maier-Leibnitz Zentrum (MLZ), D-85748 Garching, Germany}
 \affiliation{Institut für Kristallographie, RWTH Aachen, D-52056 Aachen, Germany}

\author{T. E. Schrader}
 \affiliation{J\"ulich Centre for Neutron Science (JCNS) at Heinz Maier-Leibnitz Zentrum (MLZ), D-85748 Garching, Germany}

\author{C. Pfleiderer}
 \affiliation{Physik-Department, Technische Universit\"{a}t M\"{u}nchen, D-85748 Garching, Germany}
 \affiliation{Zentrum f\"ur QuantumEngineering (ZQE), Technische Universit\"at M\"unchen, D-85748 Garching, Germany}
 \affiliation{Munich Center for Quantum Science and Technology (MCQST), Technische Universit\"at M\"unchen, D-85748 Garching, Germany}

\date{\today}

\begin{abstract}
We report a comprehensive study of single crystals of the noncentrosymmetric rare-earth compound EuPtSi$_{3}$ grown by the optical floating-zone technique. Measurements of the magnetization, ac susceptibility, and specific heat consistently establish antiferromagnetic order of localized Eu$^{2+}$ moments below the N\'{e}el temperature $T_{\mathrm{N}} = 17$~K, followed by a second magnetic transition at $T_{\mathrm{N1}} = 16$~K. For a magnetic field along the easy $[001]$ axis, the magnetic phase diagram is composed of these two phases. For fields applied in the magnetically hard basal plane, two additional phases emerge under magnetic field, where the in-plane anisotropy is weak with $[100]$ being the hardest axis. At the phase transitions, the magnetic properties exhibit hysteresis and discrepancies between differential and ac susceptibility, suggesting slow reorientation processes of mesoscale magnetic textures. Consistently, powder and single-crystal neutron diffraction in zero field identify magnetic textures that are modulated on a length scale of the order of $100~\textrm{\AA}$, most likely in the form of N\'{e}el-type antiferromagnetic cycloids.
\end{abstract}

\maketitle

\section{Motivation}

In rare-earth intermetallic compounds the interplay of localized moments with itinerant electrons may give rise to a plethora of phenomena including complex forms of magnetic order~\cite{1978_Morin_PhysLettA, 1990_Morin_Book, 2019_Kurumaji_Science}, quantum critical behavior~\cite{1979_Steglich_PhysRevLett, 2007_Lohneysen_RevModPhys, 2008_Gegenwart_NatPhys}, and unconventional superconductivity~\cite{2004_Bauer_PhysRevLett, 2005_Kimura_PhysRevLett, 2009_Pfleiderer_RevModPhys}. While cerium-based compounds are most commonly studied, due to the comparably simple electronic configuration of Ce$^{3+}$ with a single 4f electron, europium-based compounds offer an intriguing alternative. For europium, the oxidation state Eu$^{2+}$ is rather stable due to the electron configuration 4f$^{7}$ with a half-filled f shell, resulting in a quenched orbital momentum and very weak spin--orbit coupling for a rare-earth compound. In combination with crystal structures lacking inversion symmetry, in which Dzyaloshinskii--Moriya interactions potentially are leading-order in spin--orbit coupling~\cite{1957_Dzialoshinskii_SovPhysJETP, 1960_Moriya_PhysRev, 1964_Dzyaloshinskii_SovPhysJETP}, long-wavelength modulated states may emerge in europium-based compounds in analogy to transition-metal compounds~\cite{1976_Ishikawa_SolidStateCommun, 2009_Muhlbauer_Science, 2012_Seki_Science, 2012_Togawa_PhysRevLett, 2015_Kezsmarki_NatMater, 2015_Tokunaga_NatCommun}.

A candidate for such behavior is EuPtSi$_{3}$, belonging to a class of europium-based compounds crystallizing in the noncentrosymmetric tetragonal space group $I4mm$~\cite{2012_Kumar_JPhysCondensMatter, 2012_Kaczorowski_SolidStateCommun, 2013_Goetsch_PhysRevB, 2014_Albedah_JAlloyCompd, 2014_Maurya_JPhysCondensMatter, 2015_Bednarchuk_ActaPhysPolA, 2015_Bednarchuk_JAlloyCompd, 2015_Bednarchuk_JAlloyCompda, 2016_Maurya_JMagnMagnMater, 2016_Fabreges_PhysRevB, 2017_Kakihana_JAlloyCompd, 2018_Utsumi_PhysRevB, 2019_Muthu_JPhysSocJpn, 2021_Utsumi_ElectronStruct}. In a seminal study, Kumar and colleagues investigated single-crystal platelets grown from a tin solution~\cite{2010_Kumar_PhysRevB}. Despite superconducting inclusions of the solvent obscuring the intrinsic low-temperature transport properties, the authors established that magnetism in EuPtSi$_{3}$ is carried by local Eu$^{2+}$ moments and that $[001]$ is the direction of the magnetic easy axis. Signatures at 17~K and 16~K were attributed to two subsequent phase transitions. Analyzing the size of the specific heat anomaly and $^{151}$Eu M\"{o}ssbauer spectra, the authors concluded that as a function of decreasing temperatures a paramagnetic state, an incommensurate amplitude-modulated antiferromagnetic state, and a commensurate equal-moment antiferromagnetic state are observed. Electronic structure calculations accounted for the easy axis and the size of the magnetic moments, suggesting that the magnetic structure of EuPtSi$_{3}$ is incommensurate or noncollinear~\cite{2011_Pan_PhysicaB}. So far, however, the available data were scarce, in particular for fields in the basal plane, and microscopic information on the magnetic structure was lacking. Moreover, aspects such as the in-plane anisotropy or the dependence of the magnetic properties on the temperature and field history have not been addressed.

The present study reports the growth of a phase-pure single crystal of EuPtSi$_{3}$ by means of the optical floating-zone technique, thus avoiding the issues associated with high-temperature solution growth. Magnetic phase diagrams for the major crystallographic axes are inferred from comprehensive measurements of the magnetization, ac susceptibility, and specific heat. For magnetic field parallel to the easy $[001]$ axis, two long-range ordered phases are observed. For a magnetic fields applied in the hard basal plane, namely $[100]$ and $[110]$, two additional phase pockets emerge at low temperature at elevated fields. The in-plane anisotropy is weak with $[100]$ representing the hardest axis. Hysteresis as well as discrepancies between differential and ac susceptibility suggest the formation of complex mesoscale magnetic textures. Consistently, powder and single-crystal neutron diffraction in zero magnetic field indicate a long-wavelength modulation of local antiferromagnetic order, in contrast to Ref.~\cite{2010_Kumar_PhysRevB}.

\section{Experimental Methods}

For our study, a single crystal of EuPtSi$_{3}$ was grown by means of the optical floating-zone technique using an ultra-high vacuum compatible preparation chain~\cite{2016_Bauer_RevSciInstrum}. This procedure contrasts previous studies of related compounds crystallizing in space group $I4mm$, in which poly-crystalline specimens were typically prepared from the constituents by means of melting or sintering followed by an annealing process~\cite{2009_Bauer_PhysRevB, 2012_Kaczorowski_SolidStateCommun, 2013_Goetsch_PhysRevB, 2014_Albedah_JAlloyCompd, 2014_Smidman_PhysRevB} or small single crystals were grown from high-temperature solutions using tin or indium as solvent~\cite{2010_Kumar_PhysRevB, 2012_Kumar_JPhysCondensMatter, 2014_Maurya_JPhysCondensMatter, 2015_Bednarchuk_JAlloyCompd, 2016_Maurya_JMagnMagnMater, 2017_Kakihana_JAlloyCompd, 2019_Muthu_JPhysSocJpn}. While the growth from solution circumvents the challenges arising due to the high vapor pressure of europium efficiently, at least for EuPtSi$_{3}$ inclusions of the solvent result in a superconducting transition in electrical transport measurements, potentially masking the intrinsic low-temperature behavior~\cite{2010_Kumar_PhysRevB}.

\begin{figure}
	\includegraphics[width=1.0\linewidth]{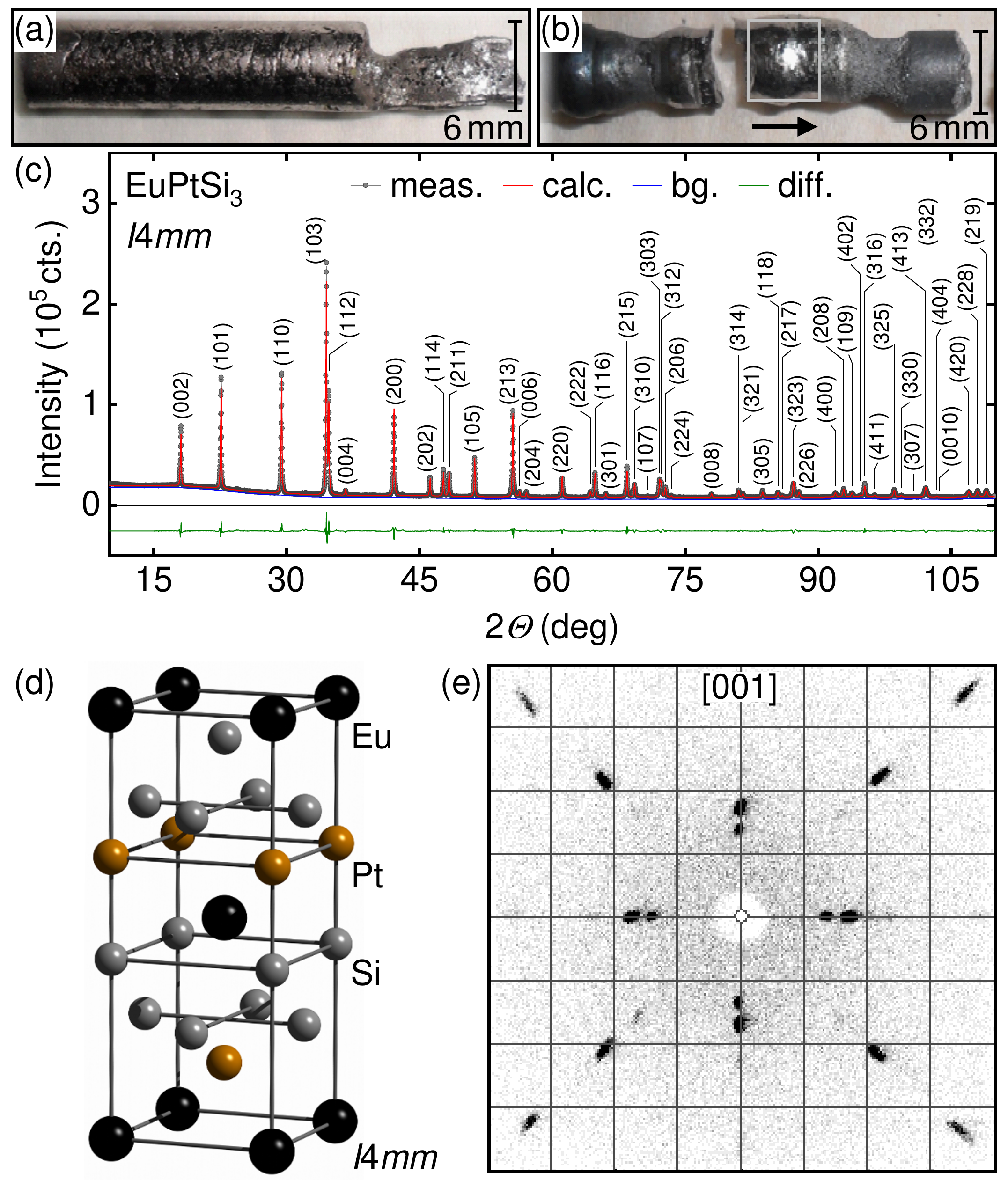}
	\caption{\label{figure01}Metallurgical analysis of EuPtSi$_{3}$. (a)~Poly-crystalline feed rod. (b)~Float-zoned ingot. The black arrow marks the growth direction. The central part yielding single-crystalline material is marked by the light gray box. (c)~X-ray powder diffraction. Taking into account the background (bg.) of the setup, measured data (meas.) and a calculated Rietveld refinement (calc.) using the space group $I4mm$ are in excellent agreement. (d)~Schematic depiction of the tetragonal unit cell. (e)~X-ray Laue diffraction pattern from a single-crystal sample showing the fourfold symmetry of the $[001]$ axis.}
\end{figure}

Such inclusions of the solvent are avoided when growing single crystals from the melt as described in the following. Using our preparation chain as optimized for intermetallic compounds, all furnaces were pumped to ultra-high vacuum prior to each step of the growth process, before an inert atmosphere was applied in the form of 6N argon additionally passed through a point-of-use gas purifier~\cite{2016_Bauer_RevSciInstrum}. Since europium is highly reactive to oxygen and moisture at ambient conditions, as a first step in an argon glovebox stoichiometric amounts of high-purity europium (4N)~\cite{url_Ames} and platinum (4N) were weighed in for the synthesis of Eu$_{5}$Pt$_{4}$~\cite{2010_Okamoto_Book}. Using a bespoke metal bellows load-lock, the elements were transferred into the horizontal cold boat furnace and melted by means of radio-frequency induction. Next, the resulting ingot, the additional platinum, and the silicon (6N) required for stoichiometric EuPtSi$_{3}$ were loaded into the rod casting furnace in which a poly-crystalline feed rod with a diameter of 6~mm and a length of ${\sim}25$~mm was cast, as shown in Fig.~\ref{figure01}(a)~\cite{2016_Bauer_RevSciInstruma}. 

In total, four such rods were prepared and used for two growth attempts in the image furnace~\cite{2011_Neubauer_RevSciInstrum}. The growth rates for the first and second attempt were 5~mm/h and 1~mm/h, respectively. In both cases, seed and feed rod were counter-rotating at 6~rpm under an inert argon atmosphere at a pressure of 5~bar. Despite evaporation losses, as observed by depositions on the inside of the image furnace presumably by the europium, stable growth conditions were obtained in both cases. The float-zoned ingots possess a shiny metallic surface without visible oxidization, as depicted in Fig.~\ref{figure01}(b) for the first growth.

The structural properties of powder prepared from the first float-zoned ingot were studied on a STOE STADI P X-ray diffractometer with copper $K_{\alpha1}$ radiation. As shown in Fig.~\ref{figure01}(c), the measured diffractogram is in excellent agreement with phase-pure EuPtSi$_{3}$ crystallizing in the noncentrosymmetric tetragonal BaNiSn$_{3}$-type structure with space group $I4mm$. A Rietveld refinement with site occupancies fixed to their nominal values yields lattice constants $a = 4.286~\textrm{\AA}$ and $c = 9.795~\textrm{\AA}$, as summarized in Tab.~\ref{table1}. These values are in acceptable agreement with those reported by Kumar and colleagues ($a = 4.2660~\textrm{\AA}$, $c = 9.8768~\textrm{\AA}$), where discrepancies may hint toward the incorporation of tin into the structure in Ref.~\cite{2010_Kumar_PhysRevB}. The crystal structure of EuPtSi$_{3}$ is illustrated in Fig.~\ref{figure01}(d), visualizing the breaking of inversion symmetry due to the distinct stacking order of the platinum and silicon layers.

Using X-ray Laue diffraction, a large single-crystal grain of a few millimeters in all dimensions was identified in the float-zoned ingot from the first growth attempt, where Fig.~\ref{figure01}(e) shows a typical Laue pattern of the fourfold $[001]$ axis. The second growth attempt using the slower growth rate did not yield single-crystalline sections of the ingot. A cube with an edge length of 2~mm and surfaces perpendicular to $[001]$, $[110]$, and $[1\bar{1}0]$ was cut from the single-crystal grain using a wire saw. Measurements of the bulk properties were carried out in a Quantum Design physical property measurement system. The magnetization was measured with the standard extraction technique of the ACMS-II option. The ac susceptibility was measured with an excitation frequency of 911~Hz at an excitation amplitude of 1~mT. The specific heat was measured using a quasi-adiabatic large heat pulse technique, where typical pulses had a size of 30\% of the temperature at the start of the pulse~\cite{2013_Bauer_PhysRevLett}.

The magnetic structure of EuPtSi$_{3}$ was studied using the high-resolution powder diffractometer SPODI~\cite{2012_Hoelzel_NuclInstrumMethodsPhysResA, 2015_Hoelzel_JLarge-ScaleResFacil} and the 4-circle single-crystal diffractometer HEiDi~\cite{2015_Meven_JLarge-ScaleResFacil}, both at the Heinz Maier-Leibnitz Zentrum~(MLZ). On SPODI, measurements were carried out in Debye--Scherrer geometry at an incident neutron wavelength of $1.55~\textrm{\AA}$. The sample consisted of ground float-zoned material with a grain size of less than $20~\mu\mathrm{m}$. Due to the high absorption cross section of europium, the powder was filled into an 0.5~mm wide gap between the walls of a hollow cylinder made of aluminum with a wall thickness of 0.2~mm and an outer diameter of 20~mm. The cylinder was mounted into a top-loading closed-cycle cryostat using helium as exchange gas. Diffraction data were collected between 3.8~K and 300~K and corrected for geometrical aberrations and the curvature of the Debye--Scherrer rings.

Full-profile Rietveld refinements of the diffraction data were carried out using pseudo-Voigt peak profiles in the FullProf program package~\cite{1993_Rodriguez-Carvajal_PhysicaB}. The background contribution was determined in terms of a linear interpolation of selected data points in nonoverlapping regions. The following parameters were varied during the fitting: scale factor, zero-angular shift, profile shape parameters, resolution (Caglioti) parameters, asymmetry and lattice parameters, as well as fractional coordinates of atoms and their displacement parameters. Since the site occupations deviated only slightly during fitting, they were fixed to their nominal values in order to suppress correlations with the displacement parameters.

On HEiDi, the single-crystal cube used for the measurements of the bulk properties was studied by means of neutrons at an incident wavelength of $1.1~\textrm{\AA}$. While cold neutrons would be favorable in terms of high resolution at small scattering angles, the absorption cross section of europium requires the use of faster neutrons in order to achieve sufficient intensity. Rocking scans on various reflections revealed Gaussian profiles with a full width at half maximum just above the experimental resolution of ${\sim}0.3^{\circ}$, confirming the single-crystalline nature and high crystalline quality of the sample with a small mosaic spread of about $0.2^{\circ}$. Using a closed-cycle cryostat, a set of specific reflections was studied in the temperature range between 2~K and 20~K. Owing the high absorption of the sample, the limited coverage of reciprocal space does not allow for a structure refinement.

As the magnetic properties of EuPtSi$_{3}$ depend on the temperature and field history, the following measurement protocols are distinguished. For data as a function of temperature, only one protocol was used. Prior to each measurement, the sample was cooled to 2~K from high temperatures well above $T_{\mathrm{N}}$ in zero magnetic field at a rate of about 5~K/min. Next, the desired field value was applied before magnetization and ac susceptibility data were collected alternately while increasing the sample temperature at a rate of 0.3~K/min. For temperatures above 30~K, the rate was increased to 1~K/min. This protocol is referred to as zero-field cooling (zfc).

For data as a function of field, the sample was first cooled from high temperatures well above $T_{\mathrm{N}}$ to the desired temperature in zero magnetic field. In the subsequent field cycle, the field was changed in steps of typically 100~mT. At each field value, the magnetization and subsequently the ac susceptibility were measured, before the next field value was approached at a rate of 10~mT/s. Five branches are distinguished: (i)~$0~\mathrm{T} \rightarrow +14~\mathrm{T}$, (ii)~$+14~\mathrm{T} \rightarrow 0~\mathrm{T}$, (iii)~$0~\mathrm{T} \rightarrow -14~\mathrm{T}$, (ii*)~$-14~\mathrm{T} \rightarrow 0~\mathrm{T}$, and (iii*)~$0~\mathrm{T} \rightarrow +14~\mathrm{T}$. Branch (i) corresponds to the situation after zero-field cooling for data as a function of temperature. Branches (ii) and (ii*) are measured under decreasing absolute field values starting in high fields. Branches (iii) and (iii*) are measured under increasing absolute field values starting from zero field, but coming from the field-polarized state in the opposite field direction. This equivalence has been checked for several temperatures (not shown).

Representing a key result of our study, magnetic phase diagrams are inferred for fields applied along $[001]$, $[100]$, and $[110]$. In measurements as a function of magnetic field, up to four signatures are observed and labeled in increasing field strength $H_{1}$, $H_{2}$, $H_{3}$, and $H_{4}$. Signatures in measurements as a function of temperature are labeled $T_{1}$, $T_{2}$, $T_{3}$, and $T_{4}$. For clarity, transition fields and temperatures are denoted with the same number when belonging to the same transition line in the phase diagram. Furthermore, the transition temperatures observed in zero magnetic field, $T_{\mathrm{N}1} = 16$~K and $T_{\mathrm{N}} = 17$~K, correspond to the origin points of the transition lines (1) and (4), namely $T_{\mathrm{N}1} = T_{1}(H=0)$ and $T_{\mathrm{N}} = T_{4}(H=0)$~\footnote{In Ref.~\cite{2010_Kumar_PhysRevB}, the two phase transition at zero-field were referred to as $T_{\mathrm{N}2} = 16$~K and $T_{\mathrm{N}1} = 17$~K. Unfortunately, the change of nomenclature was unavoidable in order to provide a consistent naming scheme in the present study.}. In fact, the positions of the transitions in phase space depend not only on the temperature and magnetic field strength, but also on the field direction. For convenience, these dependences are omitted in the nomenclature. Although this naming scheme may lead to peculiarities when considering single measurements only, it proofs to be intuitive when taking into account the magnetic phase diagrams as a whole.

\section{Experimental results}

The presentation of the experimental results starts in Sec.~\ref{overview} with the zero-field susceptibility and an account of the anisotropy of the magnetization at low temperatures. Next, for field parallel to $[001]$, in Sec.~\ref{easy} magnetization and susceptibility are shown as a function of field and temperature, complemented by the specific heat and the magnetic contribution to the entropy. Following the same organizational structure, in Sec.~\ref{hard} magnetization and susceptibility as a function of field and temperature are shown for fields parallel to $[100]$ and $[110]$. As one of the key results of this study, in Sec.~\ref{diagram} magnetic phase diagrams are presented for the three major crystallographic directions. In Sec.~\ref{neutron} the results of the powder and single-crystal neutron diffraction experiments are delineated, before the magnetic structure is discussed in Sec.~\ref{structure}.

\subsection{Magnetic key characteristics and anisotropies}
\label{overview}

\begin{figure}
	\includegraphics[width=1.0\linewidth]{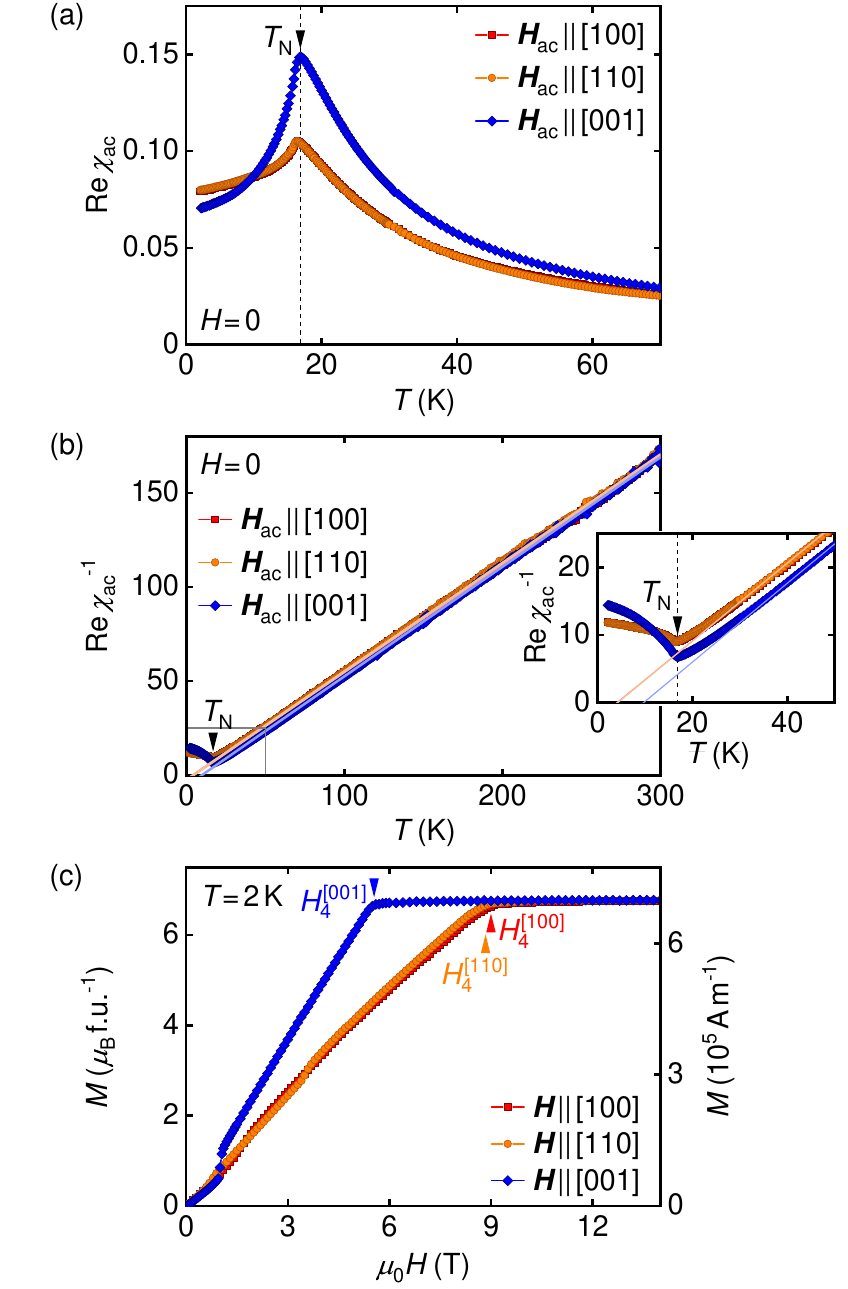}
	\caption{\label{figure02}Magnetic properties of EuPtSi$_{3}$. (a)~Temperature dependence of the ac susceptibility in zero static field for excitation fields along the major crystallographic directions. (b)~Inverse susceptibility as a function of temperature. The solid lines are linear fits indicating Curie--Weiss-like behavior. Inset: Close-up view of low temperatures. (c)~Magnetization as a function of increasing field after initial zero-field cooling at low temperature.}
\end{figure}

Shown in Fig.~\ref{figure02}(a) is the temperature dependence of the ac susceptibility for excitation fields parallel to the major crystallographic axes $[100]$, $[110]$, and $[001]$. As a function of decreasing temperature, a Curie--Weiss-like increase of the susceptibility is observed until a distinct maximum at $T_{\mathrm{N}} = 17$~K marks the onset of magnetic order. The ac susceptibility curves for excitation fields applied within the tetragonal basal plane, namely $[100]$ (red symbols) and $[110]$ (orange symbols), are almost identical and their absolute values are lower than for excitation fields along $[001]$ (blue symbols). The crossing of the curves below $T_{\mathrm{N}}$ is associated with the complex magnetic properties of EuPtSi$_{3}$, as will become clear further below.

The behavior in the paramagnetic state above $T_{\mathrm{N}}$ may be analyzed in terms of the temperature dependence of the inverse ac susceptibility, shown in Fig.~\ref{figure02}(b). Excellent agreement between the measured data and linear Curie--Weiss fits (solid lines) is observed up to room temperature. From the slope of these fits, fluctuating moments of $\mu_{\mathrm{eff}} = (7.8\pm0.1)~\mu_{\mathrm{B}}\,\mathrm{f.u.}^{-1}$ are inferred for all three directions, consistent with the free-ion value of Eu$^{2+}$, $\mu_{\mathrm{eff}} = 7.9~\mu_{\mathrm{B}}\,\mathrm{f.u.}^{-1}$. Curie--Weiss temperatures of Curie--Weiss temperatures of $(4.1\pm0.6)$~K, $(4.2\pm0.5)$~K, and $(9.7\pm0.9)$~K are inferred for excitation fields along $[100]$, $[110]$, and $[001]$, consistent with Ref.~\cite{2010_Kumar_PhysRevB}. These values are positive and distinctly smaller than $T_{\mathrm{N}}$, contrasting the expectations for both simple antiferromagnetism and ferromagnetism. Furthermore, as shown in the inset of Fig.~\ref{figure02}(b), the measured susceptibility starts to deviate from the linear fits already well above $T_{\mathrm{N}}$. In combination with the neutron scattering data indicating a long-wavelength modulation of the magnetic structure, these observations imply that contributions beyond simple exchange interactions, for instance in the form of Dzyaloshinskii--Moriya interactions, are essential for the description of the magnetic properties of EuPtSi$_{3}$ already on the mean-field level.

The field dependence of the magnetization at low temperatures, shown in Fig.~\ref{figure02}(c), provides further insight into the general character of the magnetism in EuPtSi$_{3}$. As a function of increasing field, the magnetization increases monotonically until kinks, referred to as $H_{4}$, mark the onset of the field-polarized state. At 2~K, value of $\mu_{0}H_{4}^{[100]} = 9.0$~T, $\mu_{0}H_{4}^{[110]} = 8.8$~T, and $\mu_{0}H_{4}^{[001]} = 5.5$~T are obtained for fields applied parallel to the major crystallographic axes. In agreement with the ac susceptibility, these values indicate an easy $[001]$ axis and a hard basal plane with a small but finite in-plane anisotropy and $[100]$ as the hardest axis. At high fields, the magnetization essentially saturates at $\mu_{s} = 6.8~\mu_{\mathrm{B}}\,\mathrm{f.u.}^{-1}$ independent of the crystallographic direction, consistent with the free-ion value of Eu$^{2+}$, $\mu_{s} = 7~\mu_{\mathrm{B}}~\mathrm{f.u.}^{-1}$. The good agreement of both $\mu_{\mathrm{eff}}$ and $\mu_{s}$ with their respective free-ion values indicates that the magnetism in EuPtSi$_{3}$ is described well in terms of localized Eu$^{2+}$ moments. These observations stand in contrast to the reduced values reported in Ref.~\cite{2010_Kumar_PhysRevB} which were attributed to up to 8\% of tin inclusions in the sample. At fields below $H_{4}$, changes of slope in the magnetization are characteristic of phase transitions of the magnetic order as addressed in the following.

\subsection{Properties for field along the easy $[001]$ axis}
\label{easy}

\begin{figure}
\includegraphics[width=1.0\linewidth]{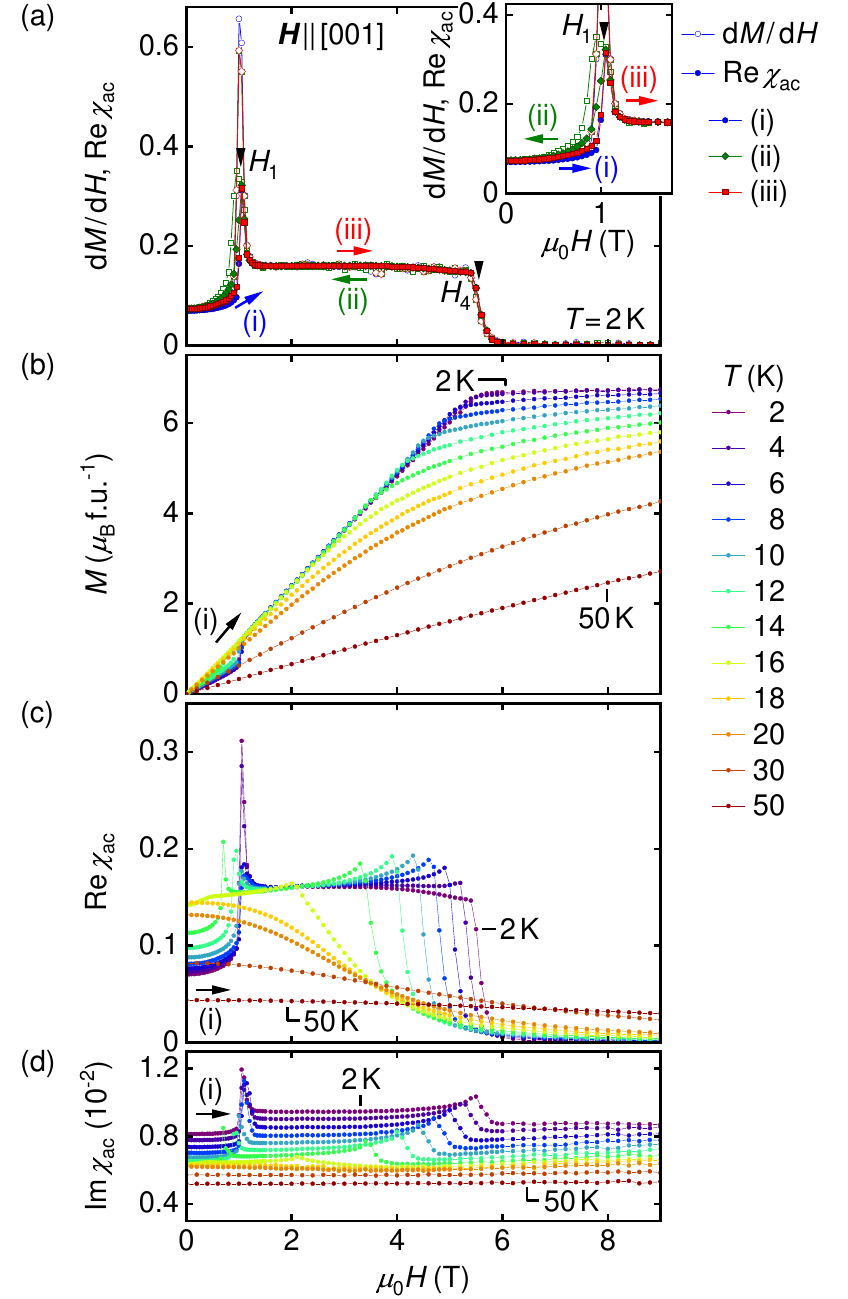}
	\caption{\label{figure03}Field dependence of the magnetic properties of EuPtSi$_{3}$ for field parallel to $[001]$. (a)~Differential susceptibility calculated from the measured magnetization, $\mathrm{d}M/\mathrm{d}H$ (open symbols), and real part of the ac susceptibility, $\mathrm{Re}\,\chi_{\mathrm{ac}}$ (solid symbols), at 2~K for different field histories. Inset: Enlarged view of the transition at low fields. \mbox{(b)--(d)}~Magnetization as well as real and imaginary part of the ac susceptibility for different temperatures.}
\end{figure}

For magnetic field along the easy $[001]$ axis, as shown in Fig.~\ref{figure03}(a), the susceptibility as a function of field at low temperature exhibits three regimes of essentially constant values. Starting at zero field with a value of 0.07, a distinct maximum at 1.0~T, referred to as $H_{1}$, is followed by a plateau at a susceptibility value of 0.16. Under further increasing field, a point of inflection at 5.5~T defines the transition to the field-polarized state, referred to as $H_{4}$. At high fields, the susceptibility exhibits a vanishingly small value, reflecting the saturation of the magnetization as a key characteristic of local-moment magnetism.

The differential susceptibility calculated from the measured magnetization, $\mathrm{d}M/\mathrm{d}H$ (open symbols), is tracked by the real part of the ac susceptibility, $\mathrm{Re}\,\chi_{\mathrm{ac}}$ (solid symbols), with exception of the transition regime around $H_{1}$, where the maximum in $\mathrm{d}M/\mathrm{d}H$ is more pronounced. Such discrepancies are abundant in materials with mesoscale magnetic textures and a characteristic of a very slow response of the system, suggesting that the transition at $H_{1}$ may involve the reorientation of large magnetic domains~\cite{2012_Bauer_PhysRevB, 2017_Bauer_PhysRevB}. Consistent with this conjecture, different field histories, labeled (i) through (iii), result in very similar behavior with only minor changes of the shape of the anomaly at $H_{1}$.

The evolution of $H_{1}$ and $H_{4}$ for increasing temperatures is illustrated in Figs.~\ref{figure03}(b), \ref{figure03}(c), and \ref{figure03}(d), where the magnetization as well as the real and imaginary part of the ac susceptibility are shown for increasing fields after zero-field cooling. Most prominently, with increasing temperature both $H_{1}$ and $H_{4}$ shift to smaller field values and have vanished at 18~K. An upturn develops in the susceptibility just before the transition to the field-polarized regime that is, however, not associated with an additional phase transition. The imaginary part of the ac susceptibility, depicted in Fig.~\ref{figure03}(d), is small and its field dependence tracks the real part qualitatively. Therefore, it can not be excluded that the imaginary part is due to a tiny phase shift well below 1~deg, which would imply that it represents an erroneous systematic contribution originating in the real part of the ac susceptibility. The putative offset of the absolute values, increasing with decreasing temperature, is due to eddy currents in the metallic sample and thus also not intrinsic.

\begin{figure}
	\includegraphics[width=1.0\linewidth]{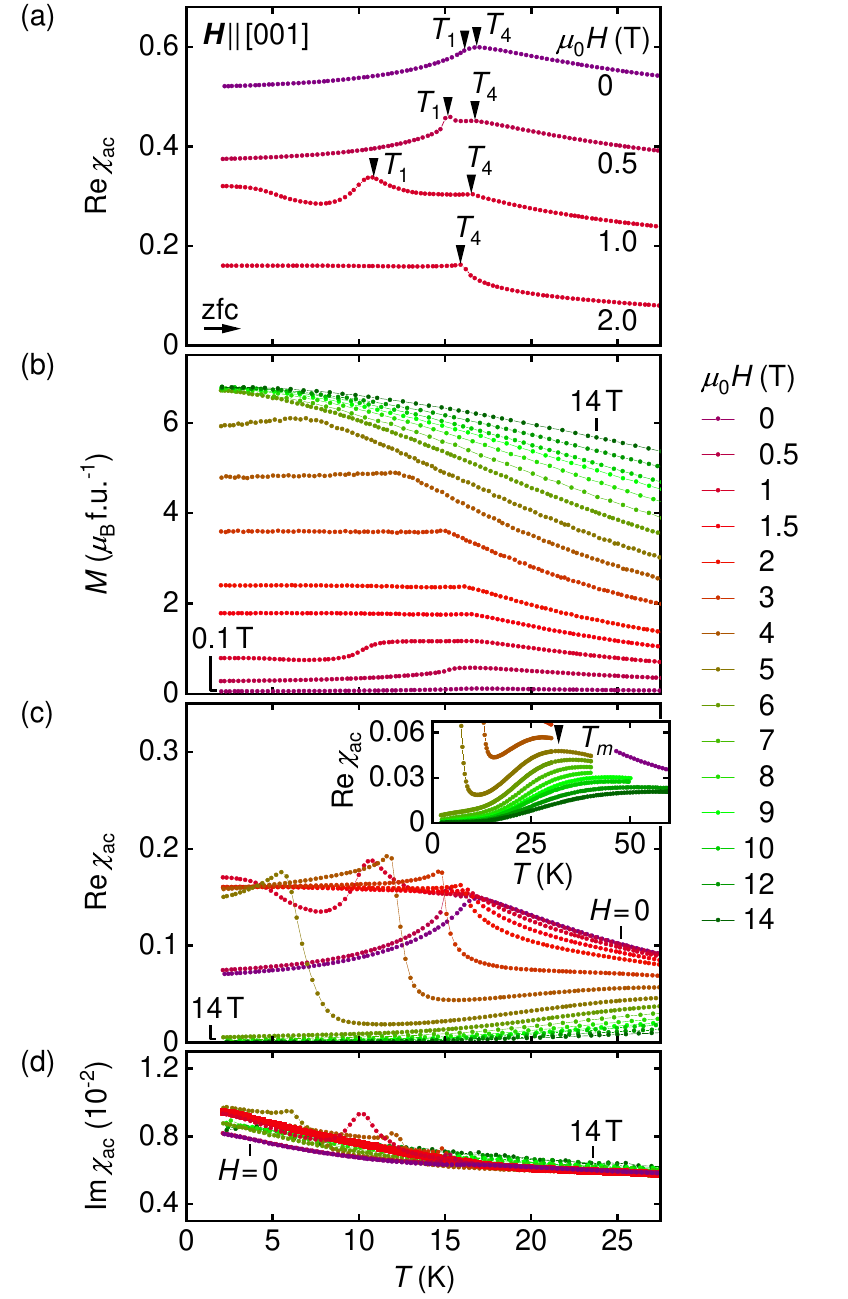}
	\caption{\label{figure04}Temperature dependence of the magnetic properties of EuPtSi$_{3}$ for field parallel to $[001]$. (a)~Real part of the ac susceptibility for selected fields measured after zero-field cooling (zfc). Data are offset for clarity. \mbox{(b)--(d)}~Magnetization as well as real and imaginary part of the ac susceptibility for a large number of fields up to 14~T. Inset: Real part of the ac susceptibility at high fields.}
\end{figure}

As a function of temperature, the real part of the ac susceptibility exhibits up to two transitions, as shown in Fig.~\ref{figure04}(a) for four field values with typical characteristics. Under decreasing temperature in zero magnetic field, the susceptibility shows a Curie--Weiss-like dependence until a kink at $T_{4}$ marks the onset of long-range magnetic order. Below $T_{4}$, the susceptibility is essentially constant until it decreases again below a second anomaly, marked $T_{1}$. The zero-field values of the transition temperatures $T_{1}$ and $T_{4}$ are also referred to as $T_{\mathrm{N}1} = T_{1}(H=0)$ and $T_{\mathrm{N}} = T_{4}(H=0)$, respectively. The presence of two independent transitions is more pronounced in small applied fields. In a field of 1~T, the susceptibility exhibits a complex shape due to the weak temperature dependence of $H_{1}$ at low temperatures. For larger fields, only the kink at $T_{4}$ remains. The transition temperatures and fields inferred from the field and temperature dependence of the susceptibility are in excellent agreement.

The evolution of the magnetic properties for a large number of field values is illustrated in Figs.~\ref{figure04}(b) through \ref{figure04}(d). With increasing field, the transition at $T_{4}$ shifts to lower temperatures until having vanished at fields of 6~T and above. Additionally, as shown in the inset of Fig.~\ref{figure04}(c), in high fields a shallow maximum in the real part of the susceptibility, referred to as $T_{m}$, is observed at temperatures well above $T_{\mathrm{N}}$. The maximum shifts to higher temperatures with increasing field and marks the crossover between the field-polarized regime at low temperatures and high fields and the paramagnetic regime at high temperatures and low fields~\cite{1997_Thessieu_JPhysCondensMatter, 2010_Bauer_PhysRevB}.

\begin{figure}
	\includegraphics[width=1.0\linewidth]{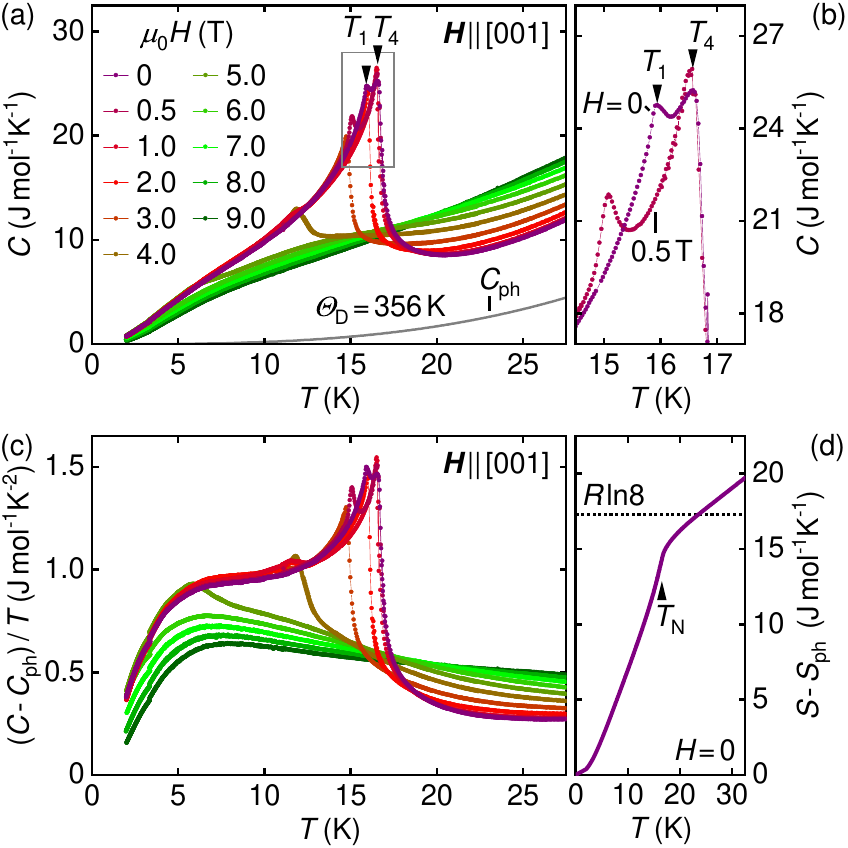}
	\caption{\label{figure05}Temperature dependence of the specific heat of EuPtSi$_{3}$ for field parallel to $[001]$. (a)~Specific heat for different magnetic fields. The gray solid line indicates the phonon contribution, $C_{\mathrm{ph}}$, using a Debye temperature $\mathit{\Theta}_{\mathrm{D}} = 356$~K. (b)~Close-up view of the transition regime. (c)~Nonphonon contribution to the specific heat divided by temperature as a function of temperature. (d)~Estimate of the entropy contribution not related to phonons.}
\end{figure}

Further information on the nature of the phase transitions may be inferred from the specific heat, depicted in Fig.~\ref{figure05}(a). In zero field, the specific heat at low temperature is dominated by a pronounced lambda anomaly at $T_{\mathrm{N}}$, characteristic of a second-order phase transition. Closer inspection reveals that the anomaly exhibits fine structure in the form a two narrowly separated maxima. As highlighted in Fig.~\ref{figure05}(b), under small applied fields along $[001]$ the low-temperature maximum distinctly shifts to lower temperatures and vanishes in fields exceeding 1~T, tracking $T_{1}$ as observed in the ac susceptibility. The high-temperature maximum associated with the $\lambda$ anomaly shifts to lower temperatures at higher fields and tracks $T_{4}$. Thus, the transition at $T_{4}$ features a substantial reduction of the entropy characteristic of a second-order transition even under large magnetic fields. In contrast, the anomaly at $T_{1}$ exhibits a rather symmetric shape and is associated with a small change of entropy only, hinting toward a first-order character of the transition.

Plotting $C/T$ as a function of $T^{2}$ (not shown) and fitting a straight line for $T > T_{\mathrm{N}}$, we determine a Debye temperature $\mathit{\Theta}_{\mathrm{D}} = 356$~K. The resulting phonon contribution with $C_{\mathrm{ph}} \propto T^{3}$ describes the behavior at intermediate temperatures reasonably well. Consistent with the assumptions of the Debye model, the fit starts to deviate from the measured specific heat for high temperatures, $T > \mathit{\Theta}_{\mathrm{D}}/10$, providing an excellent cross-check as demonstrated in detail for MnSi and related compounds in Refs.~\cite{2010_Bauer_PhysRevB, 2013_Bauer_PhysRevLett}. When subtracting the phonon contribution from the measured data, the remaining specific heat may be attributed to electronic and magnetic excitations. This contribution divided by temperature, $(C - C_{\mathrm{ph}})/T$, is shown in Fig.~\ref{figure05}(c).

For $T > T_{\mathrm{N}}$, a constant value of $0.27~\mathrm{J}\,\mathrm{mol}^{-1}\mathrm{K}^{-2}$ is observed in zero magnetic field. This contribution may be interpreted as electronic specific heat scaling linear in temperature with a relatively large Sommerfeld coefficient that is characteristic of heavy-fermion behavior. However, the slope at low temperatures suggests that $(C - C_{\mathrm{ph}})/T$ may become vanishingly small at temperatures well below 2~K. This discrepancy may hint toward an unusual breakdown of electronic correlations at low temperatures or the presence of additional effects at higher temperatures that cause a putative enhancement of the Sommerfeld coefficient. These additional complexities prevent us from separating electronic and magnetic contributions to the specific heat. Still, the numerical integration of $(C - C_{\mathrm{ph}})/T$ as a function of temperature provides an estimate of the associated entropies, as shown for zero magnetic field in Fig.~\ref{figure05}(d).

For Eu$^{2+}$, the quenched orbital momentum, $L = 0$, implies a total momentum $J = S = 7/2$ corresponds to the spin momentum. Consequently, a total magnetic entropy $S_{\mathrm{mag}} = R\ln8$ is expected at high temperatures. Consistent with the results of Kumar and colleagues, who used a measurement on poly-crystalline LaPtSi$_{3}$ for the subtraction of the phonon contribution, we obtain a value of $0.8R\ln8$ around $T_{\mathrm{N}}$. Possible explanations for this reduced value include magneto-elastic coupling, resulting in an overestimate of the phonon contribution, or magnetic correlations in the paramagnetic state above $T_{\mathrm{N}}$. Also, the entropy contribution not related to phonons clearly exceeds $R\ln8$ for temperatures well above $T_{\mathrm{N}}$, which may be attributed to electronic excitations as described above.

In previous studies, information on the magnetic structure of EuPtSi$_{3}$ was inferred from the jump of the specific heat at the magnetic phase transition. Following these considerations, the size of the anomaly would allow to distinguish between a magnetic structure with moments of the same size on all sites (equal-moment, EM) and moments of periodically varying size (amplitude-modulated, AM)~\cite{1991_Blanco_PhysRevB, 2010_Kumar_PhysRevB}. For $J = 7/2$, one expects $\Delta C_{\mathrm{EM}} = (63/26)R = 20.15~\mathrm{J}\,\mathrm{mol}^{-1}\mathrm{K}^{-1}$ and $\Delta C_{\mathrm{AM}} = \frac{2}{3}C_{\mathrm{EM}} = 13.43~\mathrm{J}\,\mathrm{mol}^{-1}\mathrm{K}^{-1}$. Kumar and colleagues estimated $\Delta C = 14.6~\mathrm{J}\,\mathrm{mol}^{-1}\mathrm{K}^{-1}$ and interpreted this value as a sign for an amplitude-modulated magnetic structure. 

In this context, we note that the experimental value in Ref~\cite{2010_Kumar_PhysRevB} was larger than the calculated one, although tin inclusions may be expected to reduce the size of the specific heat anomaly in analogy to the magnetic moment in saturation. More importantly, however, the size of the jump as inferred from the specific heat just at $T_{\mathrm{N}}$ may be erroneous. Instead, the plateau well above the onset of magnetic order in $(C-C_{\mathrm{ph}})/T$ provides a reliable point of reference, permitting us to determine the size of the jump unambiguously. In zero field, we find $\Delta C_{\mathrm{el}}/T_{c} = (1.46 - 0.27)~\mathrm{J}\,\mathrm{mol}^{-1}\mathrm{K}^{-2}$ and $T_{\mathrm{N}} = 17$~K, translating to $\Delta C = 20.2~\mathrm{J}\,\mathrm{mol}^{-1}\mathrm{K}^{-1}$. This value is in excellent agreement with a magnetic structure composed of equal moments.

\subsection{Properties for fields in the hard basal plane}
\label{hard}

\begin{figure}
	\includegraphics[width=1.0\linewidth]{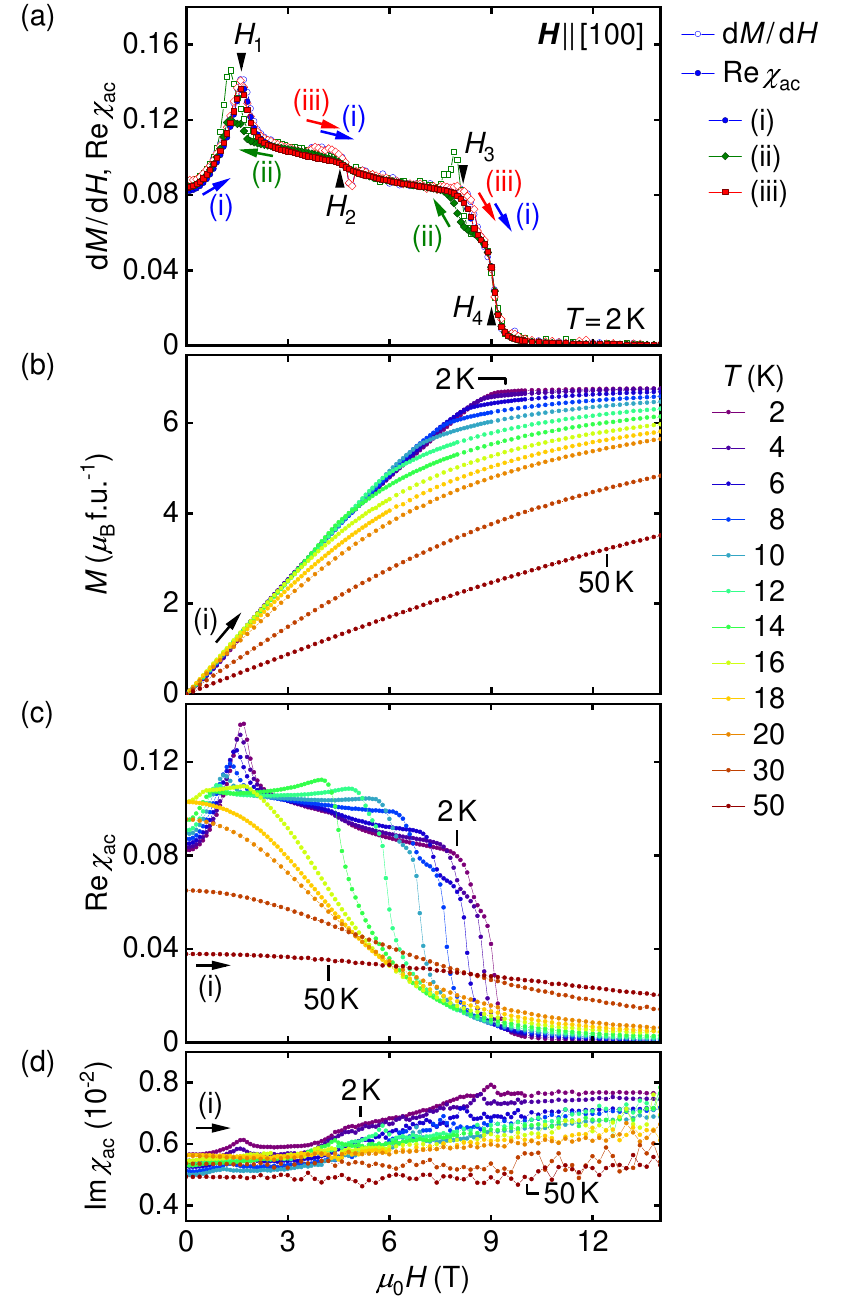}
	\caption{\label{figure06}Field dependence of the magnetic properties of EuPtSi$_{3}$ for field parallel to $[100]$. (a)~Differential susceptibility calculated from the measured magnetization, $\mathrm{d}M/\mathrm{d}H$ (open symbols), and real part of the ac susceptibility, $\mathrm{Re}\,\chi_{\mathrm{ac}}$ (solid symbols), at 2~K for different field histories. \mbox{(b)--(d)}~Magnetization as well as real and imaginary part of the ac susceptibility for different temperatures.}
\end{figure}

When applying the magnetic field in the hard basal plane, the magnetic properties are qualitatively reminiscent of the situation for field parallel to the easy $[001]$ axis. However, closer inspection reveals more complex behavior. As shown in Fig.~\ref{figure06}(a) for field along $[100]$, the differential and the ac susceptibility exhibit four distinct anomalies as a function of increasing field instead of two. Namely, the maximum at $H_{1}$ is followed by a weak kink, denoted at $H_{2}$, and a shoulder, denoted $H_{3}$, before the field-polarized regime is entered above the point of inflection at $H_{4}$.

Essentially identical behavior is observed for both histories measured under increasing field, (i) and (iii), while decreasing field, history (ii), reveals pronounced hysteresis at $H_{1}$, from $H_{1}$ to $H_{2}$, and around $H_{3}$. For the construction of the magnetic phase diagrams, the transition $H_{3}$ is defined as the center of the hysteresis. Discrepancies between the differential susceptibility, $\mathrm{d}M/\mathrm{d}H$, and the real part of the ac susceptibility, $\mathrm{Re}\,\chi_{\mathrm{ac}}$, accompany the phase transitions at $H_{1}$, $H_{2}$, and $H_{3}$, being most pronounced under decreasing field. Consistent with the results for field applied along the easy $[001]$ axis, both the history dependence and the discrepancies between differential and ac susceptibility suggest the presence of mesoscale magnetic textures and the importance of slow reorientation processes in EuPtSi$_{3}$.

With increasing temperature, the transition fields shift to lower values, as illustrated in Figs.~\ref{figure06}(b) through \ref{figure06}(d) where data are shown for increasing fields after zero-field cooling. The signatures characteristic of $H_{2}$ and $H_{3}$ vanish at lower temperatures than $H_{1}$ and $H_{4}$, enclosing distinct phase pockets in the magnetic phase diagram, as highlighted below.

\begin{figure}
	\includegraphics[width=1.0\linewidth]{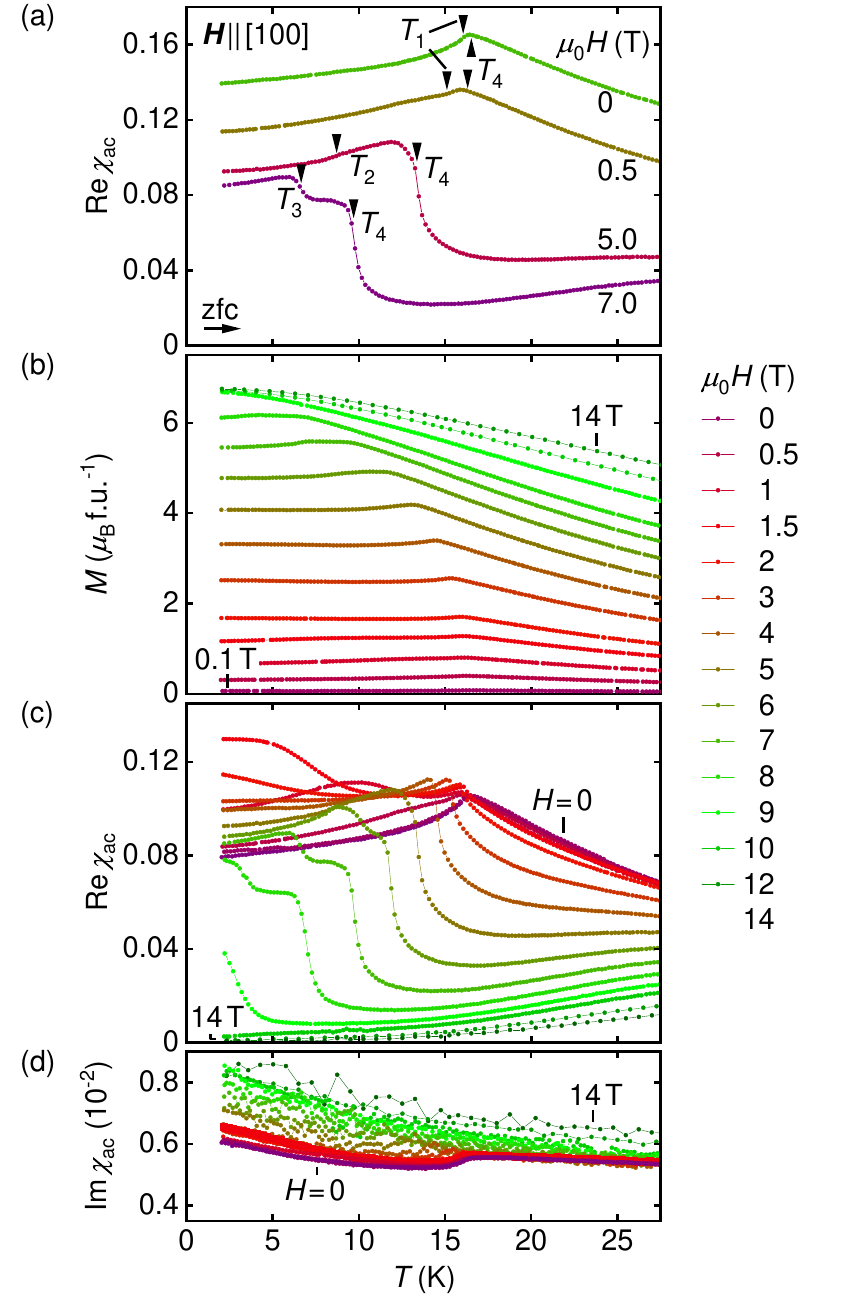}
	\caption{\label{figure07}Temperature dependence of the magnetic properties of EuPtSi$_{3}$ for field parallel to $[100]$. (a)~Real part of the ac susceptibility for selected fields measured after zero-field cooling (zfc). Data are offset for clarity. \mbox{(b)--(d)}~Magnetization as well as real and imaginary part of the ac susceptibility for a large number of fields up to 14~T.}
\end{figure}

As a function of temperature, four signatures may be distinguished, as shown in Fig.~\ref{figure07}(a) for four field values with typical characteristics. In analogy to the corresponding anomalies in the field dependence of the susceptibility, these temperatures are referred to as $T_{1}$, $T_{2}$, $T_{3}$, and $T_{4}$. Their evolution for a large number of field values is illustrated in Figs.~\ref{figure07}(b) through \ref{figure07}(d). Akin to the behavior for field parallel to $[001]$, a shallow maximum is observed in the ac susceptibility at temperatures well above $T_{\mathrm{N}}$ and associated with the crossover between the field-polarized and the paramagnetic regime (not shown).

\begin{figure}
	\includegraphics[width=1.0\linewidth]{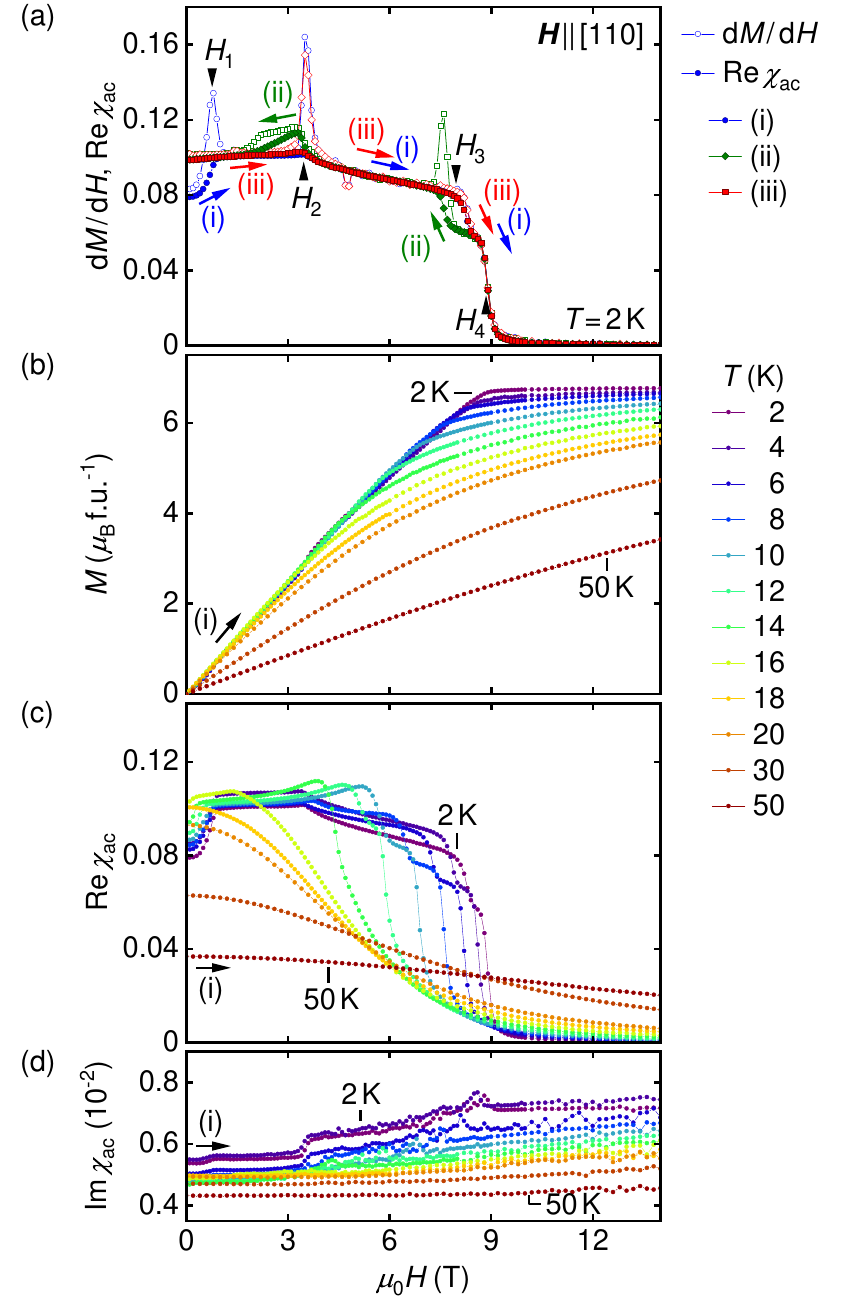}
	\caption{\label{figure08}Field dependence of the magnetic properties of EuPtSi$_{3}$ for field parallel to $[110]$. (a)~Differential susceptibility calculated from the measured magnetization, $\mathrm{d}M/\mathrm{d}H$ (open symbols), and real part of the ac susceptibility, $\mathrm{Re}\,\chi_{\mathrm{ac}}$ (solid symbols), at 2~K for different field histories. \mbox{(b)--(d)}~Magnetization as well as real and imaginary part of the ac susceptibility for different temperatures.}
\end{figure}

\begin{figure}
	\includegraphics[width=1.0\linewidth]{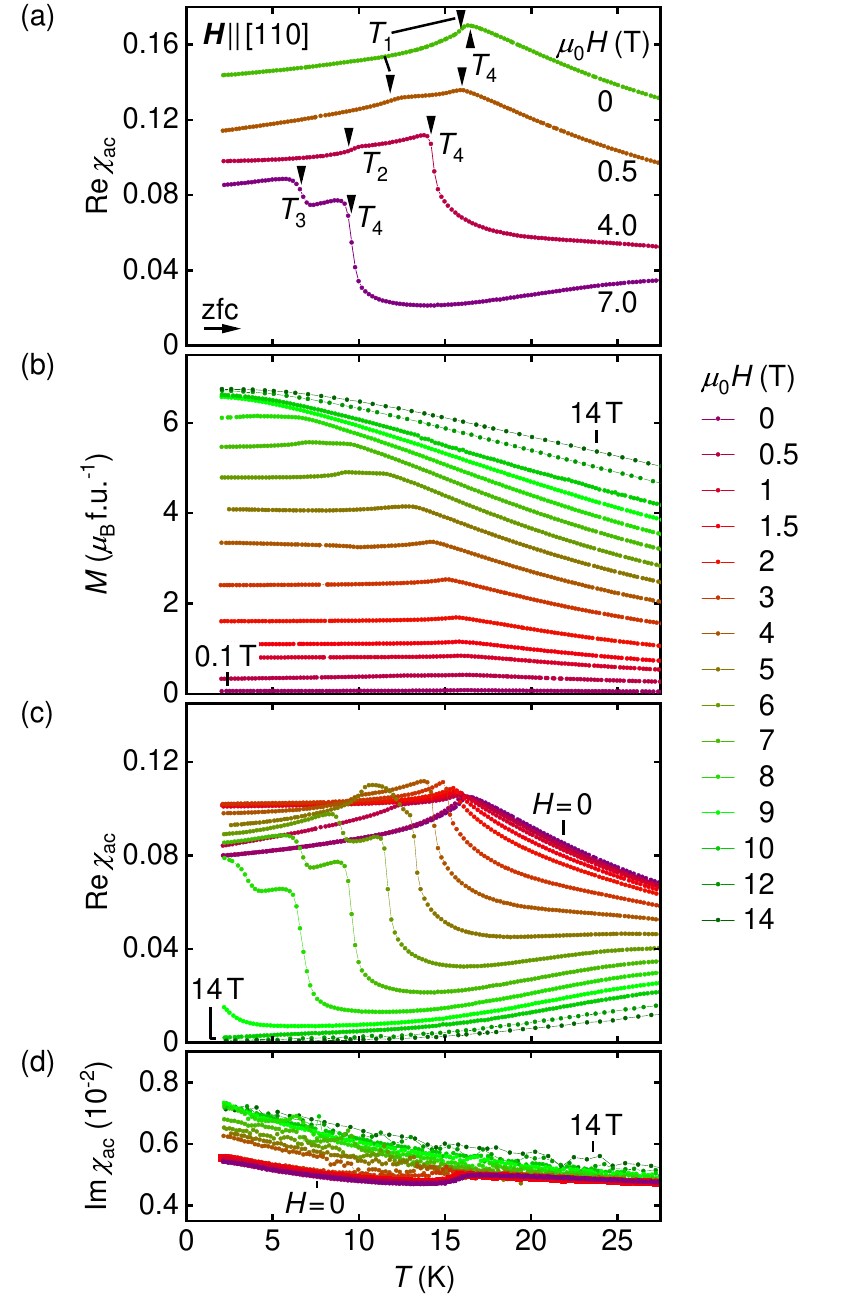}
	\caption{\label{figure09}Temperature dependence of the magnetic properties of EuPtSi$_{3}$ for field parallel to $[110]$. (a)~Real part of the ac susceptibility for selected fields measured after zero-field cooling (zfc). Data are offset for clarity. \mbox{(b)--(d)}~Magnetization as well as real and imaginary part of the ac susceptibility for a large number of fields up to 14~T.}
\end{figure}

The behavior for field along $[110]$, shown in Fig.~\ref{figure08}(a), is highly reminiscent of the situation for field along $[100]$ with four distinct anomalies being observed as a function of increasing field. Both the hysteresis between different field histories and the discrepancy between differential and ac susceptibility are more pronounced than for field parallel to $[100]$, as elaborated on in the following.

The signatures associated with $H_{1}$ are observed only after zero-field cooling, referred to as history (i), in contrast to the other field directions studied. Even for this field history, the maximum in the differential susceptibility at $H_{1}$ is not tracked by the ac susceptibility. Such a dependence may hint at a long-wavelength modulated magnetic texture for which the projection of the propagation direction into the basal plane is oriented along one of the $\langle110\rangle$ axes, in analogy to the helical-to-conical transition for magnetic field along an easy axis in cubic chiral magnets~\cite{2017_Bauer_PhysRevB}. At $H_{2}$ a sharp peak in the differential susceptibility under increasing field, histories (i) and (iii), compares with a kink in the ac susceptibility that resembles the data for field along $[100]$. Under decreasing field, history (ii), a broad asymmetric maximum is observed at slightly lower field values in both differential and ac susceptibility. The hysteretic loop around $H_{3}$ is wider than for field along $[100]$ and under decreasing field, history (ii), at its low-field boundary a peak is observed in the differential susceptibility that is not tracked by the ac susceptibility.

The evolution of the anomalies with increasing temperature, shown in Figs.~\ref{figure08}(b) through \ref{figure08}(d), as well as the magnetic properties as a function of temperature, shown in Fig.~\ref{figure09}, are highly reminiscent of the evolution for field parallel to $[100]$. Consistent with the data as a function of field, dependence on the temperature and field history is observed in particular in scans at fields below $H_{1}$ (not shown). All things considered, however, the magnetic phase diagrams for fields in the tetragonal basal plane turn out to be essentially isotropic with a slightly harder $[100]$ axis.

\subsection{Magnetic phase diagrams}
\label{diagram}

\begin{figure}
	\includegraphics[width=1.0\linewidth]{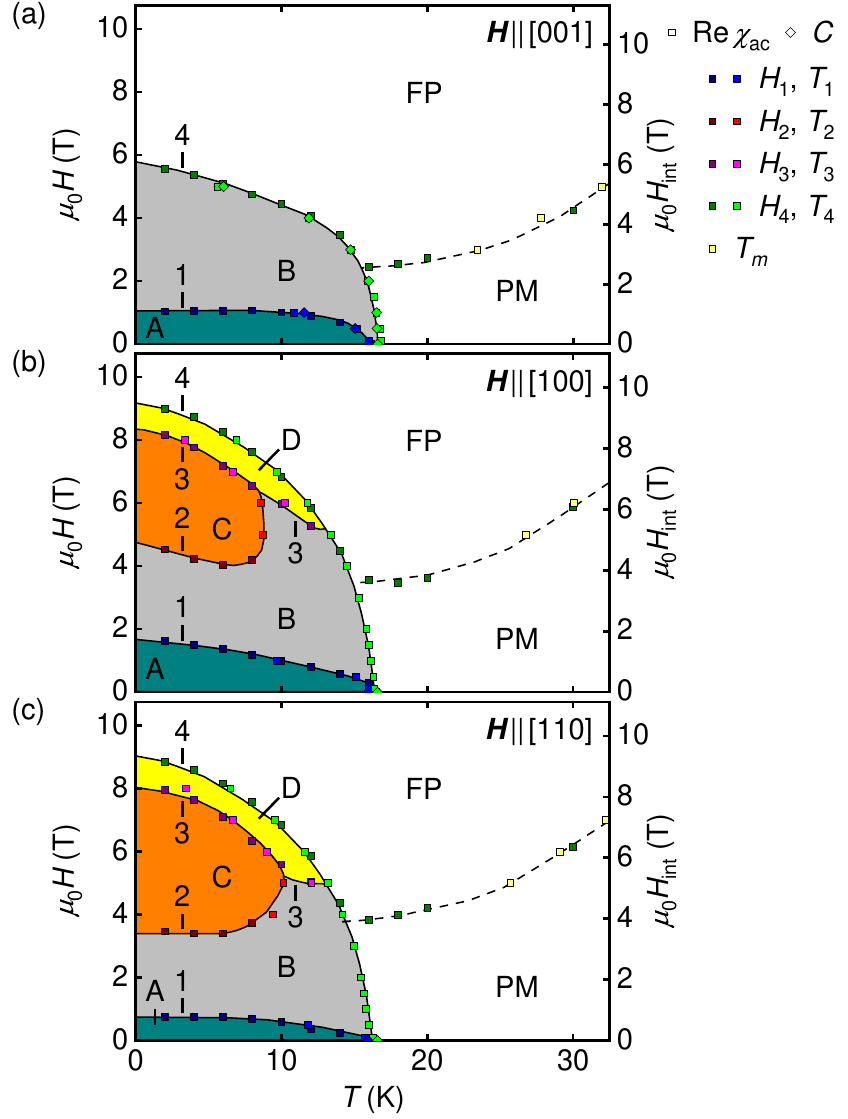}
	\caption{\label{figure10}Magnetic phase diagrams of EuPtSi$_{3}$ for field along $[001]$, $[100]$, and $[110]$ after zero-field cooling. Data inferred from measurements as a function of field and temperature are shown in dark and light colors, respectively. Paramagnetic (PM) and field-polarized (FP) regimes as well as four ordered phases, labeled A through D, are distinguished. Arabic numbers denote transition lines as defined by the transition fields and temperatures denoted with that number.}
\end{figure}

The magnetic phase diagrams of EuPtSi$_{3}$ for the three field directions studied are shown in Fig.~\ref{figure10}. The ordinate on the right-hand side displays an estimate of the internal field values $H_{\mathrm{int}}$ when correcting for demagnetization effects~\cite{1998_Aharoni_JApplPhys, 2016_Bauer_Book}. Taking together the data for all field directions, the magnetic phase diagrams feature four different phases, labeled A through D, as well as the paramagnetic state~(PM) at high temperatures and a field-polarized regime~(FP) at high fields. For zero magnetic field and decreasing temperature, the paramagnetic state undergoes a transition to phase B at $T_{\mathrm{N}} = T_{4}(H=0) = 17$~K, followed by a transition to phase A at $T_{\mathrm{N2}} = T_{1}(H=0) = 16$~K.

Under increasing magnetic field along $[001]$, the transition between phases A and B occurs at $H_{1}$, followed by a transition from phase B to the field polarized-regime at $H_{4}$. The transition fields emanate from $T_{\mathrm{N2}}$ and $T_{\mathrm{N}}$, respectively, and increase monotonically with decreasing temperature. This phase diagram is consistent with Ref.~\cite{2010_Kumar_PhysRevB}, apart from two noticeable exceptions. First, the presence of phase B in zero field was not addressed. Second, for the transition line denoted (1) a slightly convex shape was reported, implying reentrant behavior. No such behavior is observed in our study.

For magnetic fields in the basal plane, two additional states emerge, labeled C and D, before the field-polarized regime is entered. Hysteresis under increasing and decreasing fields is observed at the transitions between phases B and C, denoted transition line (2), as well as C and D, denoted transition line (3). For field along $[110]$, the hysteresis is more pronounced and, perhaps most notably, signatures associated with phase A are observed after zero-field cooling only. These hysteretic effects as well as the discrepancies between the differential and the ac susceptibility consistently suggest the presence of mesoscale textures of a long-wavelength modulated magnetic state with phase transitions that are governed by slow reorientation processes. The zero-temperature extrapolation of the transition fields $H_{1}$, $H_{2}$, $H_{3}$, and $H_{4}$ along $[110]$ are a few hundred millitesla smaller than for field along $[100]$, characteristic of weak in-plane anisotropy featuring $[100]$ as the hardest axis.

\subsection{Neutron diffraction}
\label{neutron}

\begin{figure}
	\includegraphics[width=1.0\linewidth]{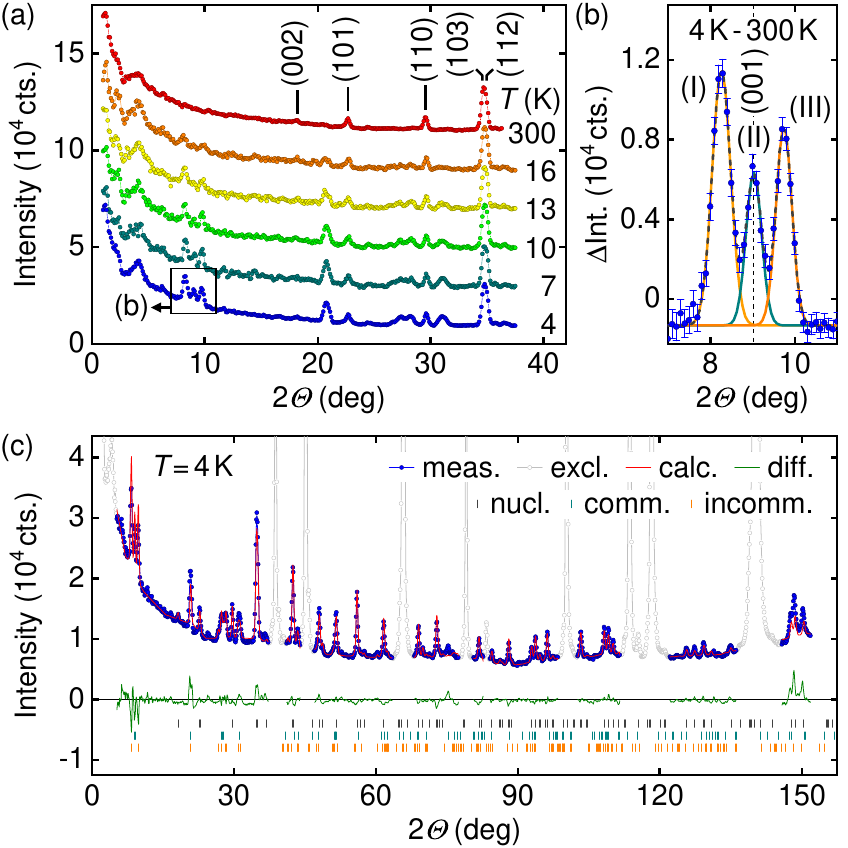}
	\caption{\label{figure11}Neutron powder diffraction on EuPtSi$_{3}$. (a)~Diffraction patterns for different temperatures. Below the onset of magnetic order additional maxima emerge. Data are offset for clarity. (b)~Enlarged view of the triplet structure around $2\mathit{\Theta} = 9~\mathrm{deg}$ where an estimate of the pure magnetic contribution is obtained by subtraction of high-temperature data. Solid lines are Gaussian fits, the dashed line represents the sum of the three Gaussians. (c)~Rietveld refinement of low-temperature data. Measured data (meas.), calculations (calc.), and their difference (diff.) are distinguished. Angles dominated by scattering from the sample environment were excluded (excl.). Short vertical lines indicate maxima related to the crystalline structure (nucl.) and the magnetic ordering vectors (comm., incomm.).}
\end{figure} 

Microscopic information on the magnetic structure of EuPtSi$_{3}$ may be inferred from neutron diffraction. As shown in Fig.~\ref{figure11}, the powder diffractogram at room temperature is consistent with a phase-pure sample crystallizing in space group $I4mm$. The structural parameters obtained by means of a Rietveld analysis are summarized in Table~\ref{table1}, establishing excellent agreement between values extracted from neutron and X-ray diffraction, cf.\ Fig.~\ref{figure01}(a). Refinements considering Pt--Si mixed occupations on sites Pt1, Si1, and Si2 yield an upper limit for Pt--Si antisite disorder of less than 3\%.

\begin{table*}
	\caption{\label{table1}Structural parameters of EuPtSi$_{3}$ crystallizing in the tetragonal BaNiSn$_{3}$-type structure with space group $I4mm$ (No.\ 107). The structural data were modeled for Eu1, Pt1, and Si1 occupying $2a$ sites $[0,0,z]$ and Si2 occupying $4b$ sites $[0,1/2,z]$ with Eu1 being fixed to the origin, $z = 0$. Data from neutron and X-ray diffraction are in excellent agreement. Numbers in parentheses give statistical deviations of the last significant digit.}
	\begin{ruledtabular}
		\begin{tabular}{c|cccc|cccc|cccc}
			Atomic site, & \multicolumn{4}{c}{Neutron, $T = 300$~K} & \multicolumn{4}{c}{X-ray, $T = 297$~K} & \multicolumn{4}{c}{Neutron, $T = 3.8$~K} \\
			Wyckoff & \multicolumn{4}{c}{$a = 4.2832(4)~\textrm{\AA}, c = 9.7932(7)~\textrm{\AA}$} & \multicolumn{4}{c}{$a = 4.28552(2)~\textrm{\AA}, c = 9.79466(6)~\textrm{\AA}$} & \multicolumn{4}{c}{$a = 4.2718(3)~\textrm{\AA}, c = 9.7747(8)~\textrm{\AA}$}\\
			position & $x/a$ & $y/a$ & $z/c$ & $B_{\mathrm{iso}}$(\textrm{\AA})$^{2}$ & $x/a$ & $y/a$ & $z/c$ & $B_{\mathrm{iso}}$(\textrm{\AA})$^{2}$ & $x/a$ & $y/a$ & $z/c$ & $B_{\mathrm{iso}}$(\textrm{\AA})$^{2}$\\
			\hline
			Eu1, $2a$ & 0 & 0   & 0        & 1.4(2) & 0 & 0   & 0         & 1.98(6) & 0 & 0   & 0        & 1.4(2)\\
			Pt1, $2a$ & 0 & 0   & 0.646(1) & 1.1(1) & 0 & 0   & 0.6475(2) & 1.14(3) & 0 & 0   & 0.646(1) & 0.9(1)\\
			Si1, $2a$ & 0 & 0   & 0.402(2) & 2.3(4) & 0 & 0   & 0.4009(8) & 2.4(2)  & 0 & 0   & 0.398(2) & 2.2(3)\\
			Si2, $4b$ & 0 & 1/2 & 0.262(2) & 1.2(2) & 0 & 1/2 & 0.2583(5) & 1.3(1)  & 0 & 1/2 & 0.259(2) & 1.4(1)\\
		\end{tabular}
	\end{ruledtabular}
\end{table*}

As a function of decreasing temperature, the crystal structure stays unchanged with the lattice constants decreasing by up to 0.2\%. As shown in Fig.~\ref{figure11}(a), below $T_{\mathrm{N}} = 17$~K additional intensity maxima emerge that are attributed to the onset of long-range magnetic order. The present data focus on low temperatures in zero field, corresponding to phase A in the magnetic phase diagram. In order to obtain an estimate of the pure magnetic contribution, the room-temperature diffractogram is subtracted from the low-temperature data. As highlighted in Fig.~\ref{figure11}(b), the most prominent contribution is a characteristic triplet around $2\mathit{\Theta} = 9~\mathrm{deg}$, denoted (I), (II), and (III).

The description of these maxima requires the superposition of at least two ordering vectors, notably in the conventional unit cell the commensurate wave vector with $\bm{k}_{15} = (001)$ and an incommensurate wave vector $\bm{k}_{10} = (00z)$ with $z = 1.081$~r.l.u.\ at 4~K. Therefore, the triplet may be indexed as $(002)-\bm{k}_{10}$, $(000)\pm\bm{k}_{15}$, and $(000)\pm\bm{k}_{10}$. With increasing temperature, $z$ increases to 1.088~r.l.u.\ at 13~K, corresponding to a decrease of the real-space periodicity from $121~\textrm{\AA}$ to $111~\textrm{\AA}$. As shown in Fig.~\ref{figure11}(c), a Rietveld refinement based on such antiferromagnetic order with multiple ordering vectors is in very good agreement with the diffraction pattern up to high scattering angles. Here, intervals of $2\mathit{\Theta}$ were omitted (excl.) that were dominated by scattering from the sample environment. Incommensurate order in phase A contrasts the findings of Kumar and colleagues, which proposed a commensurate equal-moment antiferromagnetic structure, based on specific heat measurements and M\"{o}ssbauer spectroscopy~\cite{2010_Kumar_PhysRevB}.

Unfortunately, the present data do not allow to determine the character of the incommensurate order unambiguously. The refinement in Fig.~\ref{figure11}(c) assumes an incommensurate spin density wave with moments along the edge of the basal plane, but calculations assuming N\'{e}el-type cycloidals and Bloch-type helices describe the data equally well (not shown). Moreover, a search for incommensurate ordering vectors was carried out in the basal plane of the tetragonal crystal structure. When assuming $\bm{k}_{1} = (y+x,0,y-x)$ or $\bm{k}_{3} = (x+y,x+y,2x)$, the fits converge at ordering vectors that correspond to $\bm{k}_{15}$ and $\bm{k}_{10}$ on the resolution of our study. Consequently, if there is a modulation within the basal plane of EuPtSi$_{3}$ in phase A, it will be either weak or its wavelength will be very long. No hints for a reduction of the space group symmetry of the crystal structure were observed, such as a loss of the centering translation.

\begin{figure}
	\includegraphics[width=1.0\linewidth]{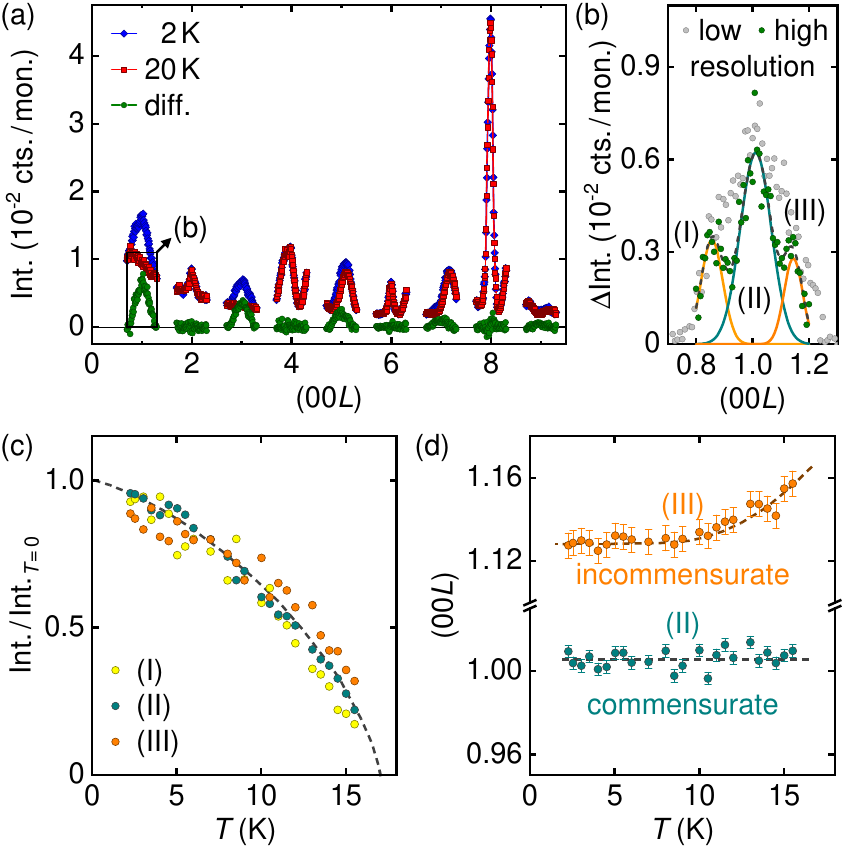}
	\caption{\label{figure12}Single-crystal neutron diffraction on EuPtSi$_{3}$. (a)~Intensity scans along $(00L)$ for different temperatures. Subtracting high-temperature data yields an estimate for the magnetic contribution at low temperatures (diff.). (b)~Enlarged view around the magnetic $(001)$ intensity maximum. A triplet structure is clearly identified in high-resolution data. Solid lines are Gaussian fits, the dashed line represents the sum of the three Gaussians. \mbox{(c),(d)}~Temperature dependence of the intensity and the position of maxima around $(001)$.}
\end{figure}

Complementary single-crystal neutron diffraction data are presented in Fig.~\ref{figure12} in the form of two scans along $(00L)$ at temperatures above and below the onset of magnetic order. In space group $I4mm$, intensity is expected only at reflections $(HKL)$ where the sum of $H$, $K$, and $L$ is even, in excellent agreement with high-temperature data in neutron and X-ray powder diffraction. At low temperatures, the single-crystal diffraction data exhibit additional maxima at positions with odd values of $L$. These maxima may be resolved best when considering the difference (diff.) of the low-temperature and high-temperature data and decrease in intensity with increasing $L$. Additional contributions independent of temperature, for instance near $(005)$ or $(007)$, appear to be spurious as they may attributed to scattering from the sample environment.

As shown in Fig.~\ref{figure12}(b), only a single broad maximum is resolved around $(001)$ on HEiDi in standard configuration (gray symbols). Placing an $\alpha_{2}$ Soller collimator with a horizontal beam divergence of $15^{\prime}$ in front of the sample allows us to resolve a triplet structure. The experimental data may be fitted with three Gaussians, yielding peak positions of 0.881~r.l.u., 1.009~r.l.u., and 1.128~r.l.u., respectively. This splitting is qualitatively consistent though larger than the splitting observed in the powder diffraction data, which may be attributed to the nonoptimal alignment of the resolution ellipsoids in our single-crystal diffraction experiment.

The integrated intensities of the three Gaussian maxima decrease simultaneously as a function of increasing temperature, as shown in Fig.~\ref{figure12}(c). This behavior clearly establishes the origin of the scattering as magnetic. However, the data available do not allow us to discriminate whether the intensity vanishes at $T_{\mathrm{N}1}$ or $T_{\mathrm{N}}$. As depicted in Fig.~\ref{figure12}(d), the position of the central maximum of the triplet remains unchanged as a function of increasing temperature, consistent with a commensurate ordering vector, while for the incommensurate maxima the separation away from $(001)$ increases from 0.128~r.l.u.\ at 2~K to 0.157~r.l.u.\ at 15.5~K.

\subsection{Magnetic structure}
\label{structure}

The crystal structure of EuPtSi$_{3}$ sets the stage for potential magnetic structures. Due to the lack of inversion symmetry, neighboring magnetic moments, $\bm{S}_{i}$ and $\bm{S}_{j}$, may be subject to the Dzyaloshinskii--Moriya interaction $\bm{D}_{ij} \cdot \left(\bm{S}_{i}\times\bm{S}_{j}\right)$ with the Dzyaloshinskii--Moriya vectors $\bm{D}_{ij}$~\cite{1957_Dzialoshinskii_SovPhysJETP, 1960_Moriya_PhysRev, 1964_Dzyaloshinskii_SovPhysJETP}. The corresponding energy is minimized when $\bm{S}_{i}$, $\bm{S}_{j}$, and $\bm{D}_{ij}$ form an orthogonal trihedron. In combination with interactions favoring collinear alignment, such as ferromagnetic or antiferromagnetic exchange interactions, the moments may form a long-wavelength modulated structure in the plane perpendicular to $\bm{D}_{ij}$.

\begin{table}
	\caption{\label{table2}Dzyaloshinskii--Moriya vectors as constrained by the crystal structure of EuPtSi$_{3}$ in space group $I4mm$. Along specific types of crystallographic directions, for an europium moment at position $\bm{R}_{i} = (0,0,0)$ low-order nearest-neighbor europium moments at positions $\bm{R}_{j}$ are considered, where $a$ and $c$ are the tetragonal lattice constants. Applying the Moriya rules~\cite{1960_Moriya_PhysRev}, allowed orientations of the Dzyaloshinskii--Moriya vector $\bm{D}_{ij}$ are inferred, thereby determining the favored type of long-wavelength modulation.}
	\begin{ruledtabular}	
		\begin{tabular}{c|c|c|c}
			direction            & $\bm{R}_{j}$    & $\bm{D}_{ij}$      & Modulation    \\
			\hline
			$\langle100\rangle$  & $(a,0,0)$       & $\pm(0,1,0)$       & N\'{e}el-type \\
			$\langle111\rangle$  & $(a/2,a/2,c/2)$ & $\pm(1,\bar{1},0)$ & N\'{e}el-type \\
			$\langle110\rangle$  & $(a,a,0)$       & $\pm(1,\bar{1},0)$ & N\'{e}el-type \\
			$\langle001\rangle$  & $(0,0,c)$       & $(0,0,0)$          & None          \\
		\end{tabular}
	\end{ruledtabular}
\end{table}

The space group $I4mm$ constrains possible Dzyaloshinskii--Moriya vectors $\bm{D}_{ij}$, as inferred from the Moriya rules~\cite{1960_Moriya_PhysRev} and summarized in Tab.~\ref{table2}. For moments neighboring along the basal edges $\langle100\rangle$, the tetragonal diagonals $\left\langle111\right\rangle$, or the basal face diagonals $\langle110\rangle$, the Dzyaloshinskii--Moriya vectors are perpendicular to the line connecting the moments. Thus, when considering long-wavelength magnetic modulations, N\'{e}el-type cycloidals are favored over Bloch-type helices. As Dzyaloshinskii--Moriya vectors with opposite sign are mapped onto each other by mirror symmetries of space group $I4mm$, the magnetic modulation is expected to inherit this achirality. For moments neighboring along $\langle001\rangle$, the Dzyaloshinskii--Moriya vector vanishes. As an additional note, the space groups $I4mm$ belongs to point group $C_{4v}$, for which magnetic vortices, so-called skyrmions, were predicted that are based on N\'{e}el-type cycloidals~\cite{1989_Bogdanov_SovPhysJETP, 1994_Bogdanov_JMagnMagnMater, 1999_Bogdanov_JMagnMagnMater}.


In combination with the results of the bulk measurements and neutron scattering, these considerations permit us to identify key characteristics of the magnetic order in EuPtSi$_{3}$. In zero magnetic field at low temperatures, namely in phase A, our data are consistent with local-moment antiferromagnetism of the europium moments that is modulated on a length scale of ${\sim}100~\textrm{\AA}$ in terms of a N\'{e}el-type cycloidal. For this conjecture, the following aspects are crucial. 

First, the crystal structure belongs to space group $I4mm$ which lacks inversion symmetry and therefore may support long-wavelength magnetic modulations stabilized by means of Dzyaloshinskii--Moriya interactions, favoring N\'{e}el-type cycloidals rather than Bloch-type helices. Second, magnetization and specific heat data indicate that the magnetism is carried by localized Eu$^{2+}$ moments with quenched orbital momentum and equal size on all sites. Third, the differential and the ac susceptibility exhibit a pronounced history dependence, in particular for field along the basal face diagonals $\langle110\rangle$, suggesting the presence of mesoscale magnetic domains. Fourth, neutron scattering is in agreement with commensurate antiferromagnetic order that is superimposed by a long-wavelength incommensurate magnetic texture propagating along $\langle001\rangle$. Fifth, taken together, the two previous aspects suggest either (i) propagation along $\langle001\rangle$ with domains featuring projections of moments parallel to $\langle110\rangle$ or (ii) propagation along directions slightly tilted away from $\langle001\rangle$ toward $\langle110\rangle$ such that the peak splitting is well below the resolution of our neutron scattering study.

\section{Conclusions}

In summary, our study reported the growth of a single crystal of EuPtSi$_{3}$, crystallizing in the noncentrosymmetric tetragonal space group $I4mm$, by means of the optical floating-zone technique. This single crystal was studied by means of magnetization, ac susceptibility, and specific heat as well as powder and single-crystal neutron diffraction. The magnetization and specific heat are characteristic of localized Eu$^{2+}$ moments. As a function of decreasing temperature, two narrowly separated magnetic phase transitions are observed at $T_{\mathrm{N}} = 17$~K and $T_{\mathrm{N2}} = 16$~K. Under applied magnetic field along the easy $[001]$ axis, two phases are observed for which the phase transition lines connect with $T_{\mathrm{N}}$ and $T_{\mathrm{N}1}$. At low temperatures, the field-polarized state is entered above 5.5~T. Under applied field in the hard basal plane, two additional phases emerge in larger fields. The in-plane anisotropy is weak, where $[100]$ represents the hardest axis, with the field-polarized state at low temperatures forming above 9.0~T and 8.8~T for field along $[100]$ and $[110]$, respectively.

Neutron scattering in zero field at low temperatures exhibits a splitting of magnetic intensity maxima, characteristic of a long-wavelength modulated antiferromagnetic texture, most likely in the form of a N\'{e}el-type cycloid. These findings highlight that the quenched orbital momentum of intermetallic europium compounds leads to relatively weak spin--orbit coupling, making them prime candidates for long-wavelength magnetic modulations, in particular when combined with crystal structures lacking inversion symmetry, thus permitting Dzyaloshinskii--Moriya interactions.

\begin{acknowledgments}
We wish to thank P.\ B\"{o}ni, G.\ Causer, F.\ Haslbeck, J.\ Leiner, S.\ Mayr, and A.\ Regnat for fruitful discussions and assistance with the experiments. Moreover, we want to thank U.\ Keiderling for test measurements at the small-angle scattering instrument V4 at the Helmholtz-Zentrum Berlin~(HZB). We also want to thank S.\ M\"{u}hlbauer and E.\ Blackburn for test measurements at the small-angle scattering instrument SANS1~\cite{2015_Heinemann_JLarge-ScaleResFacil} using a 17~T superconducting magnet and A.\ Ostermann for assistance with test measurements at the diffractometer BIODIFF~\cite{2015_Ostermann_JLarge-ScaleResFacil}, both beamlines at the Heinz Maier-Leibnitz Zentrum~(MLZ). Parts of the data were collected on HEiDi, jointly operated by RWTH Aachen University and the J\"{u}lich Centre for Neutron Science (JCNS) within the JARA cooperation. This work has been funded by the Deutsche Forschungsgemeinschaft (DFG, German Research Foundation) under TRR80 (From Electronic Correlations to Functionality, Project No.\ 107745057, Project E1) and the excellence cluster MCQST under Germany's Excellence Strategy EXC-2111 (Project No.\ 390814868). Financial support by the European Research Council (ERC) through Advanced Grants No.\ 291079 (TOPFIT) and No.\ 788031 (ExQuiSid) is gratefully acknowledged.
\end{acknowledgments}

\end{document}